\documentclass{article}
\usepackage[utf8x]{inputenc}
\usepackage{authblk}
\usepackage{url}

\title{Application of Deep Neural Networks to assess corporate Credit Rating}
\author[1]{Parisa Golbayani}
\author[1,2]{Dan Wang} 
\author[1,2]{Ionu\c{t} Florescu\footnote{Corresponding author. \texttt{ifloresc@stevens.edu}, Other authors emails: \texttt{dwang35@stevens.edu}, \texttt{pgolbaya@stevens.edu}}}
\affil[1]{Financial Engineering, School of Business, Stevens Institute of Technology, 1 Castle Point Terrace, Hoboken, NJ, 07030, USA}
\affil[2]{Hanlon Financial Laboratories , Stevens Institute of Technology, 1 Castle Point Terrace, Hoboken, NJ, 07030, USA}
\date{Running header/short title: Deep Neural Networks for corporate Credit Rating\\
Number of words: 4441\\
Number of Figures: 0 \\
Number of Tables: 10
}
\date{}

\usepackage{longtable}
\usepackage{lipsum}
\usepackage{natbib}
\usepackage{graphicx}
\usepackage{amsmath}
\usepackage{amsthm}
\usepackage{float}
\usepackage{adjustbox}
\usepackage{subcaption}
\usepackage[colorlinks=true,linkcolor=blue,citecolor = blue]{hyperref}%
\usepackage{tikz}
\usetikzlibrary{tikzmark}
\usepackage{multirow}
\usepackage{marginnote}
\usepackage{comment}

\theoremstyle{definition}

\usepackage{xargs}                      % Use more than one optional parameter in a new commands
\usepackage[colorinlistoftodos,prependcaption,textsize=small]{todonotes}
\newcommandx{\unsure}[2][1=]{\todo[linecolor=red,backgroundcolor=red!15,bordercolor=red,#1]{#2}}
\newcommandx{\change}[2][1=]{\todo[linecolor=blue,backgroundcolor=blue!25,bordercolor=blue,#1]{#2}}
\newcommandx{\info}[2][1=]{\todo[linecolor=OliveGreen,backgroundcolor=OliveGreen!25,bordercolor=OliveGreen,#1]{#2}}
\newcommandx{\improvement}[2][1=]{\todo[linecolor=Plum,backgroundcolor=Plum!25,bordercolor=Plum,#1]{#2}}
\newcommandx{\thiswillnotshow}[2][1=]{\todo[disable,#1]{#2}}

\begin{document}

\maketitle
\begin{abstract}
Recent literature implements machine learning techniques to assess corporate credit rating based on financial statement reports. In this work, we analyze the performance of four neural network architectures (MLP, CNN, CNN2D, LSTM) in predicting corporate credit rating as issued by Standard and Poor's. We analyze companies from the energy, financial and healthcare sectors in US. The goal  of the analysis is to improve application of machine learning algorithms to credit assessment. To this end, we focus on three questions. First, we investigate if the algorithms perform better when using a selected subset of features, or if it is better to allow the algorithms to select features themselves. Second, is the temporal aspect inherent in financial data important for the results obtained by a machine learning algorithm? Third, is there a particular neural network architecture that consistently outperforms others with respect to input features, sectors and holdout set? We create several case studies to answer these questions and analyze the results using ANOVA and multiple comparison testing procedure.
\end{abstract}
\textbf{ECL:}{ C45, C52, C55}

\section{Introduction}
Credit rating is an indication of the level of the risk in investing with the company. It represents the likelihood that the company pays its financial obligations on time. Standard \& Poor’s uses a two-fold analysis (Qualitative and Quantitative) to assign a credit score to a company. Qualitative analysis is based on different factors such as company strategy and economic market outlook, while quantitative analysis is only based on financial statements. However, how these analysis lead to the final credit score is still unclear \cite{smith2003neural}.

There are number of machine learning methods that have been extensively employed to forecast corporate credit ratings. The architecture of these models allows to explore complex non linear trends in data. Artificial neural network is one of the most frequently used techniques in such analysis. 
Du compares the performance of traditional Back Propagation neural network (BP) with Genetic Algorithm Back Propagation neural network (GABP) on 100 listed companies in China in 2017. GABP has been used as an alternative optimization technique to find the optimal values of neural network \cite{du2018enterprise}. Miyamoto, et al. apply neural network on 1 million Japanese SMEs \cite{miyamotopredicting}.
Huang et al. applied back propagation neural network on Taiwan and United States datasets to forecast corporate credit ratings \cite{huang2004credit}. Angelini et al. deployed two different topologies of neural networks on 76 small businesses of a bank in Italy. The studied topologies are feed forward classical neural network and feed forward neural network with ad hoc connections. The latter is a feed forward neural network where input neurons are grouped together \cite{angelini2008neural}. Nazari et al. studied classical neural network to measure bank customers credit risk in Iran \cite{nazari2013measuring}. H{\'a}jek used various neural network structures to classify US municipalities which are located in Connecticut into rating classes. They focus on Moody's rating classes as benchmark of their model \cite{hajek2011municipal}. Yu et al. studied six stage neural network ensemble model to evaluate credit scores. \citep{yu2008credit}
%At first stage, they use bagging sampling to obtain different training datasets. These datsets will be used as input data for different neural networks at the second stage. After training neural networks in stage three, 

In addition to standard neural network, some researchers focus on deep learning models in many financial fields. Deep Neural Network (DNN) is typically an Artificial Neural Network (ANN) with more than one hidden layer between the input and output layers \cite{dixon2017classification}. Deep learning models such as Convolutional Neural Networks (CNN) \cite{lecun1995convolutional} and Recurrent Neural Networks (RNN) \cite{hochreiter1997long} has been proved to significantly improve upon traditional machine learning techniques in various financial problems \cite{fu2016credit}.

Previous applications of CNN are mostly concentrated around stock market analysis \cite{liu2019anticipating,rajaa2019convolutional,ntakaris2019feature,chavan1d,dixon2017classification}.  Tsantekidis et al. use CNN on a large scale limit order book dataset to predict mid-price movements. They obtain more accurate results comparing to MLPs when trying to predict short term price changes  \cite{tsantekidis2017forecasting}. Chaven et al. propose a one-dimensional CNN to predict future patterns in stock market \cite{chavan1d}. Arratia et al. used recurrent plot (RP) to increase dimension of financial time series as an input to the CNN. Dereli et al. use CNN on text regression problem in which they predict stock return volatility from companies annual report \cite{dereli2019convolutional}. Takeuchi and Lee use DNN on times series of stock prices to find important features that can significantly effect on stock returns \cite{takeuchi2013applying}.

RNN and LSTM have been applied on many problems such as speech recognition \cite{graves2005framewise, graves2006connectionist} and handwriting recognition \cite{liwicki2007novel,graves2008unconstrained}. In finance, both RNN and LSTM have been widely used in financial time series prediction in particular in the field of stock price prediction \cite{roondiwala2017predicting,zhuge2017lstm,minami2018predicting,rather2015recurrent,rout2017forecasting}. Some researchers used historical time series to predict the stock prices \cite{rout2017forecasting,rather2015recurrent}. Others took one step further and studied the impact of the environment of the stock along with the market conditions \cite{zhuge2017lstm}. Hsieh et al. combine wavelet transforms and RNN to predict stock markets \cite{hsieh2011forecasting}. 
Xiong et al. use Long Short Term Memory (LSTM) to predict S$\&$P500 volatility \cite{xiong2015deep}. LSTM works well on a broad range of parameters such as input and output gate biases and does not require any parameter tunning \cite{hochreiter1997long,gao2016stock}.

The purpose of this study is to examine a standard neural network on US corporates belong to healthcare, financial and energy sectors. Furthermore, we study deep learning models such as CNN and LSTM to forecast corporate credit ratings. To the best of our knowledge this is the first work that uses a CNN and LSTM for predicting corporate credit ratings.

%Wojcicka uses Multi Layer Perceptron (MLP) and Radial Basis Function (RBF) for fraud detection \cite{wojcickacan}. There has been also a number of other papers on neural networks bankruptcy predictions \cite{atiya2001bankruptcy,vellido1999neural,zhang1999artificial, fu2016credit}. 

\section{Data sets studied}\label{dataset}
Real-world data are often noisy and incomplete. Therefore, the first step of any prediction problem, in particular, to credit risk assessment is to clean data such that we maintain as much meaningful information as possible.  
The data in this study was obtained from Bloomberg and Compustat. It consists of $332$ financial variables of three financial, energy and healthcare sectors in US. We gathered all financial variables available in Bloomberg and Compustat for $30$ companies in energy sector from $2010-2016$, $66$ companies in financial sector from $2000-2016$ and $59$ companies in healthcare sector from $2000-2016$. We use Standard and Poor's credit ratings as our benchmark. This study considers two types of data frames. One includes a selected subset of input features and the other one is based on all input variables. Details of these analysis are provided in section \ref{sec:selectedfeatures}. With respect to train/test datasets, we conduct random and non-random frameworks. In both cases, we split the dataset into train and holdout sets. In case of non-random selection, we form the test set by holding out one year, and the remaining data will be used as the training set. In order to have a fair assessment in our comparison between these two datasets, in random selection for these three sectors, we split the dataset  for  energy  sector  to $85\%$  training  and  $15\%$  test  set.   For  financial  and  healthcare sectors, we split the dataset to $94\%$ training and $6\%$ test set. This percentage split in data roughly corresponds to the  split based on a year in the test data and the rest of years in the training data set.

In all these cases, the holdout set does not participate in developing the model. The validation set is supposed to be used to adjust the model parameters and detect when overfitting happens. Instead of using validation set, we repeated each experiment $15$ times to assess the stability of the model. In order to address overfitting, we employed built-in regularization terms called ''Dropout'' and ''early stopping''. The idea behind ''early stopping'' is to stop training as soon as the error on the validation set is higher than it was the last time it was checked \cite{prechelt1998early}. Another important approach to reduce the error on the holdout set and to address overfitting, is to combine the predictions of many different models (Dropout) \cite{breiman2001random}. In order to create these different models, Hinton, et al. suggested to use the probability of each hidden unit. In this technique, the output of hidden unites will set to zero if their probability goes below a specific threshold. These hidden units do not participate in the forward pass and consequently in the backpropagation process. Therefore, each time an input is fed to the model, the neural network will have a different architecture \cite{hinton2012improving,krizhevsky2012imagenet}. 

\section{The Neural Network Architectures used}
In this section, we provide a broad description and specific details of the network architecture used in this study. For more details please refer to the citations of each section. 
 
\subsection{Multi-Layer Perceptron (MLP)}\label{mlp}

This is the most classical neural network architecture, also known as feed forward Artificial Neural Network. It consists of an input layer, an output layer and several hidden layers. In MLP, all the nodes are fully connected to the nodes in adjacent layers. MLP is the most prevalent network architecture for credit rating problems \cite{ahn2011corporate,huang2004credit,kumar2003forecasting, kumar2006artificial}.

Large scale neural networks are typically overfitting the training data sets. There are several parameters of the MLP such as, the number of hidden layers, the number of hidden units, the learning rate and the dropout ratio which are crucial for tuning the performance of such networks. In this study, we start by using the values of these parameter (also known as hyper parameters) published in \cite{huang2004credit}. Then, we use \textit{GridSearch} to find the best values of these parameters for our specific datasets.

% In this work we use three hidden layers. The number of hidden units in each hidden layer is 2 * number of input features + 1. In this study, we have 20 input features, thus there are 41 hidden units in each hidden layer. We study the effect of changing these two parameters on overfitting. Tables \ref{tab:1}, \ref{tab:2} and \ref{tab:3} present the numerical results obtaining when changing these parameters in MLP. 

\subsection{CNN}\label{cnn} 
CNNs are a derivative of traditional Multilayer Perceptron (MLP) neural networks. They are optimized for advanced computer vision tasks such as two dimensional pattern recognition and image classification problems \cite{driss2017comparison, lecun1995convolutional}. 

The architecture of CNN  consists of three key components: the convolution layers, the pooling layers and the fully connected layers. CNNs assume there is local connectivity between the input vector data points. They capture this connectivity by using filters with specific size and stride, panning around the entire image. Each filter moves along the input data $k$ points at a time (stride of $k$) and extracts certain features by applying a convolution operation on input tensor. When the stride value increases, the dimension of the input data decreases. In case of images, the input data is in a matrix format. Margins (edges) of images are a very different part of input than a rectangle in the middle of the image. To detect the edges of the image, generally the image is padded with 0's. When dealing with large dimensional input data (e.g., high resolution images), CNN is introducing pooling layers between subsequent convolution layers. 
The only purpose of the pooling layer is to reduce the spatial dimension of the input. The depth of the original data set will be unchanged as the pooling is independently done on each dimension.  The most common type of pooling is max pooling which computes the maximum value of data on each consecutive set of vectors in a given dimension. The combination of convolution and pooling layers captures the temporal dynamics of time series \cite{tsantekidis2017forecasting}. After the last convolution/pooling layers, a set of fully connected layers are used in the network. The ground truth target for us $y_i$ is the number of distinct classes we have for corporate credit scores while $\tilde{y_i}$ is the predicted credit score for a given observation . The error in prediction is obtained by applying the cross entropy loss function on the output layer. CNN is trained using a classic backpropagation algorithm. The backpropagation algorithm \cite{ werbos1990backpropagation} is based on minimizing the categorical cross entropy loss function.
% \dan{after read other paper, I found if people will derectly use mlp, cnn model, they either state in detail about the model, or they explain the model by words. I believe it is better to simply delete those formulas for cnn, and descrbe the logic by words}

% $$ L(W) = - \sum_{i=1}^{S} y_i \cdot \log \tilde{y_i} $$

% where $S$ is the number of unique corporate credit scores and $W$ refers to the  parameters of the CNN such that $W_i \in R^{N * D * M}$ where $N$ is the number of filters, $D$ is the dimension of each filter and $M$ is the number of input channels \cite{tsantekidis2017forecasting, kim2014convolutional}. 

% CNN also uses the so-called gradient descent backpropagation algorithm \cite{ werbos1990backpropagation} to minimize the loss function defined above. 

% $$ W^{\prime} = W − \eta \cdot \frac{\partial f}{\partial x}$$

% where $\eta$ is the learning rate and $W^{\prime}$ is the result of gradient descent step and.

Comparing to standard feed forward neural network, CNN is keeping an easier training process due to the less number of parameters and connections. It is also capable of selecting useful financial features during the training process as a result of convolution operation \cite{di2016artificial}.

In this work we implement two types of CNN; one-dimensional (CNN) and two-dimensional CNN (CNN2D). In both architectures, we use two convolution layers and two fully connected layers. The first convolution layer includes $64$ filters, the second convolution layer consists of $32$ filters, each with size $3$ that move along the input data by stride = $1$. The last two fully connected layers contain $128$ neurons each. In constructing CNN, the input data contains only one quarter (the current quarter), so the filter moves only in one direction. 
However, in CNN2D, the 4 most recent consecutive quarters are fed as input to the network at a time, so the filter moves in two directions.

\subsection{LSTM}

Corporate credit scores do not change very much from quarter to quarter. Therefore, studying changes in credit rating is much more important than treating each quarter as an independent observation. However, credit rating does not deteriorate immediately, signs of distress will typically appear in quarters before the rating is changing. This led us to include in the input variables information from the previous quarters and naturally to the Long Short Term Memory (LSTM) neural network architecture \cite{hochreiter1997long}. 

LSTM is a special type of recurrent neural networks (RNN). RNN is a type of neural network containing loops, that allow information to persist in time. One of the important advantages of RNN is the ability to connect previous temporal observations as input for the problem. In finance, this is typically crucial which explains why LSTM are popular in this area. For a more detailed description of the LSTM architecture please see \cite{heaton2016deep, dixon2017classification}. For different applications of LSTM in finance see \cite{akita2016deep, bao2017deep, chen2015lstm}.

In this study, we implement LSTM with three layers. The first layer contains LSTM cells with 32 units. The second and third layers are densely-connected neural network layers each with 128 units. We consider 4 consecutive trades as the input to LSTM. 
% \begin{remark}[CNNLSTM]
% We note that LSTM is designed to handle temporal data and does not have a convolutional layer to deal with complex input points. However, in our research both steps are important and thus we also implement a recurrent network with a convolutional layer. We denote this architecture as CNNLSTM henceforth. 
% \end{remark}

\section{Experiment results and analysis}

We are addressing three different questions in this section.
\begin{itemize}
\item \textit{First, is the performance better when using a selected subset of input features versus all input variables?} (Section \ref{sec:selectedfeatures})
\item \textit{Second, is the performance different depending on the choice of training/test datasets?}  (Section \ref{sec:crossvalidation})
\item \textit{Third, is there a particular neural network architecture that consistently outperforms others with respect to input features, sectors and holdout set?} (Section \ref{sec:compareNN})\label{questions}
\end{itemize}

\subsection{\textit{Comparing the performance of algorithms when using a selected set of features versus all input features}\label{sec:selectedfeatures}}
Rating agencies use specific financial ratios to assess corporate credit ratings \cite{wallis2019credit}. These features are considered to be good indicators of companies financial situation. In this section, we choose features that are important for companies profitability and revenue \cite{wallis2019credit, kumar2001detection}. Table \ref{tab:informative features} provides the list of these features according to the literature provided.
% Most literature on using machine learning techniques to predict corporate credit score, apply various types of feature selection techniques to find informative features \ion{\cite{}}. For example, Revenues, ROE, Dividend Yield are commonly encountered in most studies. 
\textbf{Case1:} In this section, we use these selected features as an input to the feed forward neural network to predict corporate credit ratings of three sectors (financial, energy and healthcare) in US. 

\begin{table}[H]
\caption{Financial Ratios as common informative features}
\centering  
    \begin{tabular}{ccc}
    \hline
         &Ratio \\
         \hline
         \hline
         $R_1$&Debt/EBITDA\\
         
         $R_2$&FFO/Total Debt\\
         
         $R_3$&EBITDA/Interest\\
         
         $R_4$&FFO/Interest\\
         
         $R_5$&CFO/Debt\\
         
         $R_6$&FFO/Net Profit\\
         
         $R_7$&NWC/Revenue\\
         
         $R_8$&Current Asset/Current Liabilities\\
         
         $R_9$&(FFO+Cash)/Current Liabilities\\
         
         $R_{10}$&EBITDA/Revenues\\
         
         $R_{11}$&Cash/Total Debt\\
         
         $R_{12}$&Total Debt/Tangible Net worth\\
         
         $R_{13}$&Total Debt/Revenue\\
         
         $R_{14}$&Debt/Capital\\
         
         $R_{15}$&Cash/Asset\\
         
         $R_{16}$&Total Fixed Capital/Total Fixed Assets\\
         
         $R_{17}$&Equity/Asset\\
         
         $R_{18}$&NWC/Total Assets\\
         
         $R_{19}$&Retained Earnings/Total Assets\\
         
         $R_{20}$&EBITDA/Total Assets\\
         \hline
    \end{tabular}
    
    \label{tab:informative features}
\end{table}

For the case 1 we only use standard MLP neural network architecture (Section \ref{mlp}). Similar results are obtained for the other network architectures. As in \cite{huang2004credit} we start with 41 hidden units, and use \textit{GridSearch} to find the best values of hidden units along with optimal number of hidden layers for our datasets. For future reference, the best results correspond to a 3 hidden layers architecture with the learning rate equaling $0.05$.  To address overfitting we apply cross validation, dropout and early stopping.

\begin{table}[H]
    \centering
    \caption{Cross Validation, Dropout and Early Stopping results on energy sector}
    \label{tab:3}
    \begin{tabular}{|c|c|c|}
    \hline
         Number of hidden units & Accuracy on test set & Accuracy on train set  \\
         \hline 
         \hline
         41  & 0.776 & 0.819\\
         82  & 0.808 & 0.806\\
         164  & 0.753 & 0.838\\
         \hline
    \end{tabular}
    
\end{table}

 Table \ref{tab:3} shows the optimal accuracy on the test and training set on energy sector. 

% \begin{table}[H]
%     \centering
%     \begin{tabular}{|c|c|c|}
%     \hline
%          Hyper Parameters & FFNN1 & FFNNX\\
%          \hline 
%          \hline
%          Activation function& RELU& RELU\\
%          Batch norm& True& True\\
%          Batch size& 128& 128\\
%          Dropout& 0.50& 0.75\\
%          Early Stopping& 1& 10\\
%          Learning Rate& 0.001& 0.001\\
%          Number of hidden units& 39& 143\\
%          number of hidden layers& 1& 2\\
%          Optimizer& adam & adam \\
%          \hline
%     \end{tabular}
%     \caption{GridSearch Results for FFNN and DNN}
%     \label{tab:3}
% \end{table}

% \parisa{Parisa: make the tables and results}

\textbf{Case 2:} In this scenario we use all historical financial variables available on \textit{COMPUSTAT} indiscriminately as input data set to neural networks. We also use the MLP architecture to have a direct comparison with Case 1. 

The data contains $332$ financial variables and we replace all missing points with zero. Handling missing values is important, as they can significantly affect the results of classification. To address this issue, we also implement a CNN architecture. Recall from section \ref{cnn}, that the convolutional layer in the CNN architecture is capable of discovering important features automatically. Specifically, in our case CNN adapts the weights of features to classify the respective corporate credit ratings.

\begin{table}[H]
    \centering
    \caption{Mean and Standard Deviation of accuracy based on random selection using 332 financial variables}
    \label{tab:stats_shuffle}
    \begin{tabular}{|c|c|c|c|c|}
    \hline
         & Test Mean & Test Std & Train Mean & Train Std  \\
         \hline
         \hline
         Energy & & & &\\
	 \hline
	 MLP & 0.872 & 0.010 & 0.974 & 0.009\\
	 CNN & 0.885 & 0.022 & 0.984 & 0.034\\
%	 CNN2d & 0.932 & 0.0182 & 0.993 & 0.0212\\
%	 LSTM & 0.960 & 0.0138 & 0.977 & 0.017\\

         \hline
         Financial & & & & \\
         \hline
	 MLP & 0.747 & 0.022 & 0.837 & 0.022\\
	 CNN & 0.868 & 0.012 & 0.981 & 0.007\\
%	 CNN2d & 0.885 & 0.0123 & 0.983 & 0.008 \\
%	 LSTM & 0.884 & 0.017 & 0.958 & 0.018 \\
	 \hline
         Healthcare & & & &\\
         \hline
	 MLP & 0.776 & 0.014 & 0.869 & 0.014\\
	 CNN & 0.866 & 0.0324 & 0.981 & 0.026\\
%	 CNN2d & 0.842 & 0.023 & 0.992 & 0.009\\
%	 LSTM & 0.875 & 0.017 & 0.958 & 0.016\\
	 \hline

    \end{tabular}
    
\end{table}
Table \ref{tab:stats_shuffle} compares the performance of MLP and CNN on test and training sets for energy, financial and healthcare sectors. As we can clearly see CNN provides better results. We believe this indicates that handling missing data is important and probably MLP may be improved by better processing data.\footnote{All experiments in this section are based on random selection of train/test dataset. Energy sector contains $30$ companies from 2010-2016. Healthcare and financial sectors include $59$ and $66$ companies respectively from 2000-2016. In order to have a fair assessment, we split the dataset for energy sector to $85\%$ training and $15\%$ test set. For financial and healthcare sectors, we split the dataset to $94\%$ training and $6\%$ test set. We train and test each model $15$ times to determine the stability of each model.}

% The mean and standard deviations in Table \ref{tab:stats_shuffle} shows that LSTM and CNN2D outperform CNN and MLP models. In most cases, LSTM provides a better accuracy comparing to CNN2D. 

Table \ref{tab:test_shuffle_comparison} showcase the main comparison of this section. We present the results obtained applying MLP on the test data for the energy and healthcare sectors. We compare case 1 which develops a model based only on selected features while case 2 MLP model contains all the variables. From the results we see that even using an inefficient way of handling missing data\footnote{replace all missing points with $0$, no convolutional layer}, case 2 is outperforming the results for case 1. 

\begin{table}[H]
    \centering
    \caption{Comparison of mean and standard deviation of accuracy on the test set for Energy and Healthcare sectors }
    \begin{tabular}{|c|c|c|c|c|}
    \hline
         Sector &  Case 1 mean & Case 1 std dev & Case 2 mean & Case 2 std dev \\
         \hline
         Energy & 0.808 & 0.032 & 0.872 & 0.010 \\
         \hline
         Healthcare & 0.752 & 0.025 & 0.776 & 0.014 \\
         \hline
    \end{tabular}
    
    \label{tab:test_shuffle_comparison}
\end{table}
The difference is statistically significant and imply that when using Neural Networks it is better to use all features rather than a small selected subset. The results we obtain are different than the general literature on the subject. Multiple papers are dedicated to selecting the most relevant features for the credit rating problem. Our results may also indicate that the set of best features should be viewed in the context of the particular classification/clustering method used. Specifically, a particular set of features that are important for credit rating using SVM may not be the same as the best features for a credit rating using random forests.  

\subsection{\textit{Comparing randomly selection of data with yearly selection of data}\label{sec:crossvalidation}}
When forecasting corporate credit ratings, the majority of existing literature is using a random selection of data points to select the training and testing subsets. However, in finance, most datasets have a temporal aspect to them. If one uses random selection to separate train/test data it is possible that the model is trained using the same quarter data as the test data. We believe the temporal aspect is important as quarterly credit rating is done around the same time. One would think that the existing political and economic conditions will have an effect on the ratings. For example, consider 2008 the most turbulent year in our data set. Realistically, one would like the training set to not contain any quarter from that year. The reason is that the financial statements from different companies in the same sector during that time are similar. 

To test the effect of random allocation of data between training and test datasets we create two test cases. 

\textbf{Case 3:} We run four different neural network architectures: MLP, CNN, CNN2D and LSTM for all sectors. We randomly split the dataset into a training set and a holdout set as in the previous section. As we explained in section \ref{dataset}, we split the dataset for the energy sector into an $85\%$ train and $15\%$ test sets. For financial and healthcare sectors, we split the dataset into $94\%$ training and $6\%$ test set.
% Energy sector contains data from 2010-2016. Healthcare and financial sectors contain data from 2000-2016.
% \dan{we have similar description for dataset in section2, and we also explain why the do different percentage split in section 2}\parisa{i changed it} In order to have a fair assessment, we split the dataset for the energy sector into an $85\%$ train and $15\%$ test sets. For financial and healthcare sectors, we split the dataset into $94\%$ training and $6\%$ test set. 
The goal is to compare results of this case with the ones obtained for Case 4 with the same ratio of train/test set. %This percentage split in data roughly corresponds to the  split based on a year in the test data and the rest of years in the training data set.

\textbf{Case 4:} We form the test set by selecting one year, and the train data as all the years minus the one selected. This allocation avoids having data from the same period in the test and training sets. We again apply to all sectors using MLP, CNN, CNN2D and LSTM models.
Table \ref{tab:Yearly_no_difference} provides the results of this analysis obtained from the test data. For completion we show the results for all years. For example, the 2014 row entry means that the training set contains data for all years but 2014 and test data is 2014. 
 Due to space constraints we decided not to include the results for the training data. However, we should mention that the accuracy numbers obtained for the training data are much larger than those in table \ref{tab:Yearly_no_difference} obtained for the test data. This is generally indicative of overfitting. Intuitively, this is as expected since there are no contemporaneous points in the training and test data sets and thus the performance on the test data is much worse. 

\begin{longtable}{|c|c|c|c| c|}
    
    % \centering
    % \setlength\tabcolsep{4pt}
    \hline
         Energy & CNN & CNN2D  & LSTM  &  MLP   \\
         \hline
         2010 & 0.788  & 0.835 & 0.825  & 0.769  \\
         2011 & 0.889  & 0.917 & 0.870  & 0.898   \\
         2012 & 0.830  & 0.973 & 0.929  & 0.893  \\
         2013 & 0.893  & 0.902 & 0.902  & 0.839  \\
         2014 & 0.848  & 0.830 & 0.839  & 0.813  \\
         2015 & 0.716  & 0.843 & 0.913  & 0.690  \\
         2016 & 0.543  & 0.875 & 0.938  & 0.491  \\
         \hline
         \hline
         Financial & CNN & CNN2D & LSTM  & MLP  \\
         \hline
         2000 & 0.565  & 0.770  & 0.745  & 0.503  \\
         2001 & 0.706  & 0.822  & 0.789  & 0.567  \\
         2002 & 0.801  & 0.828  & 0.806  & 0.581  \\
         2003 & 0.817  & 0.869  & 0.649  & 0.749  \\
         2004 & 0.776  & 0.890  & 0.810  & 0.711  \\
         2005 & 0.807  & 0.894  & 0.787  & 0.715  \\
         2006 & 0.699  & 0.815  & 0.731  & 0.611  \\
         2007 & 0.662  & 0.676  & 0.740  & 0.557  \\
         2008 & 0.505  & 0.631  & 0.653  & 0.486   \\
         2009 & 0.576  & 0.768  & 0.817  & 0.629   \\
         2010 & 0.808  & 0.885  & 0.836  & 0.734   \\
         2011 & 0.811  & 0.829  & 0.899  & 0.759 \\
         2012 & 0.860  & 0.856  & 0.908  & 0.790  \\
         2013 & 0.875  & 0.862  & 0.806  & 0.888 \\
         2014 & 0.803  & 0.803  & 0.849  & 0.777 \\
         2015 & 0.867  & 0.846  & 0.817  & 0.863   \\
         2016 & 0.723  & 0.935  & 0.919  & 0.752  \\
        \hline 
        \hline 
         Healthcare & CNN & CNN2D  & LSTM  & MLP  \\
         \hline 
         2000 & 0.525  & 0.803 	& 0.752		& 0.516	 \\
         2001 & 0.611  & 0.765 	& 0.788		& 0.634	 \\
         2002 & 0.690  & 0.638 	& 0.696		& 0.620	 \\
         2003 & 0.718  & 0.754 	& 0.810		& 0.620	\\
         2004 & 0.676  & 0.759 	& 0.841		& 0.572	\\
         2005 & 0.707  & 0.804 	& 0.811		& 0.660	 \\
         2006 & 0.772  & 0.771 	& 0.785		& 0.655	 \\
         2007 & 0.722  & 0.711 	& 0.792		& 0.656	 \\
         2008 & 0.704  & 0.803 	& 0.770		& 0.724	\\
         2009 & 0.816  & 0.829 	& 0.868		& 0.770 \\
         2010 & 0.732  & 0.680 	& 0.830		& 0.686	 \\
         2011 & 0.775  & 0.811 	& 0.855		& 0.794	\\
         2012 & 0.806  & 0.839 	& 0.845		& 0.838	 \\
         2013 & 0.788  & 0.785 	& 0.767		& 0.764	 \\
         2014 & 0.747  & 0.858 	& 0.876		& 0.729	 \\
         2015 & 0.806  & 0.804 	& 0.881		& 0.876	 \\
         2016 & 0.696  & 0.882 	& 0.863		& 0.702	\\
         \hline
         
    \caption{Accuracy based on Case 4}
    \label{tab:Yearly_no_difference}

\end{longtable}

To formally compare the results of the Cases 3 and 4, Table \ref{tab:stat_yearly_nodifference} presents the mean and standard deviation of percent accuracy for the test dataset. Case 4 column presents average performance across years with standard deviation in parenthesis. For Case 3 we used $15$ different random allocations, while the Case 4 uses each year as one observation.

\begin{table}[htbp]
  \centering
    \caption{Mean and Standard Deviation of percent accuracy for Cases 3 and 4}
    \label{tab:stat_yearly_nodifference}
    \begin{tabular}{|l|l|l|}
    \hline 
          & Case3 Mean (Std) & Case 4 Mean (Std) \\
    \hline 
    Energy &       &  \\
    \hline 
    mlp   & 0.8359(0.0200) & 0.7704(0.1427) \\
    cnn   & 0.8775(0.0243) & 0.7868(0.1237) \\
    lstm  & 0.9543(0.0154) & 0.888(0.0438) \\
    cnn2d & 0.9403(0.0243) & 0.8822(0.0522) \\
%    cnnlstm & 0.1425(0.0206) & 0.4887(0.1158) \\
    \hline 
    Financial &       &  \\
    \hline 
    mlp   & 0.7321(0.018) & 0.6866(0.1206) \\
    cnn   & 0.8654(0.0145) & 0.7447(0.1117) \\
    lstm  & 0.8984(0.0255) & 0.7978(0.0775) \\
    cnn2d & 0.9013(0.0275) & 0.8222(0.0776) \\
%    cnnlstm & 0.1876(0.0069) & 0.3363(0.1038) \\
    \hline 
    Healthcare &       &  \\
    \hline 
    mlp   & 0.8181(0.0242) & 0.695(0.0941) \\
    cnn   & 0.8972(0.0123) & 0.723(0.0746) \\
    lstm  & 0.8862(0.0245) & 0.8135(0.0509) \\
    cnn2d & 0.8444(0.0288) & 0.7821(0.0623) \\
%    cnnlstm & 0.1153(0.0225) & 0.4111(0.0579) \\
    \hline 
    \end{tabular}%
  \label{tab:addlabel}%
\end{table}%

There are a few observations we can draw from the results in Table \ref{tab:stat_yearly_nodifference}. It appears that a random split of data into test/training (case 3) generally produce better results than a more realistic temporal allocation (case 4). To confirm this we perform a one sided test with the following hypotheses: 
\begin{small}
\begin{align*}
   & H_0: \text{ There is no difference in results obtained for yearly versus random allocation} \\
    &H_a: \text{ The results for random allocation are better than those for yearly allocation}
\end{align*}
\end{small}
We present the $p$ values of the test in Table \ref{tab:two_sided_t_test}. From these numbers we see that, with the exception of MLP for the energy and financial sector, CNN for energy sector, every result is significantly larger when using a random allocation of data. The numbers points out the need to have a proper temporal allocation when testing machine learning algorithms developed for credit rating. A random allocation of data will significantly increase the percent accuracy.  

\begin{table}[H]
\centering
\caption{P values for a one-sided t test comparing results obtained for case 3 and case 4 }
   \label{tab:two_sided_t_test}
  
  \begin{tabular}{|c|c|c|c|}

	\hline
	
	& Energy & Financial & Healthcare \\
	
	\hline

	\hline
	
    MLP   & 0.1359 & 0.0714 & 2.84E-05 \\
    CNN   & 0.0506 & 0.0002 & 1.69E-08 \\
    LSTM  & 0.0033 & 0.0000 & 1.17E-05 \\
    CNN2D & 0.0126 & 0.0004 & 5.78E-04 \\

%	 CNNLSTM & 2.83e-10 & 5.28e-06 & 5.44e-18\\

    \hline
 
   \end{tabular}
 
\end{table}

A second observation from the numbers presented in Table \ref{tab:Yearly_no_difference} is that LSTM and CNN2d seem to outperform CNN and MLP over all three sectors. We expected this difference in results as both CNN2D and LSTM use as input the current quarter as well as the past 3 quarters, while MLP and CNN use as input only the current quarter. The next section is dedicated to discovering the best network architecture.

\subsection{\textit{Comparing the performance of neural networks}}\label{sec:compareNN}

In this section we introduce a formal statistical test that compares the performance of the network architectures considered. To our surprise, none of the machine learning articles perform this type of testing procedure. Most are using a simple t-test similar to the testing procedure in the previous section. This test is suitable for a single pair comparison but when we are performing multiple comparisons the p-value of the tests has to be suitably modified. Here we are using a multiple comparison procedure (see for example \cite{hsu1996multiple}). 

Specifically, we consider a two way ANOVA model where the response variable is the performance of the model. The first factor is the sector with 3 levels: energy, financial and healthcare. The second factor is the network architecture with 4 levels: MLP, CNN, CNN2D, and LSTM. Table \ref{tab:2-way-anova} presents the ANOVA table for this analysis. We can clearly see that the interaction term is highly significant. This implies there is no universally best network architecture and in fact each sector needs to be studied separately. 

\begin{table}[H]
    \centering
    \caption{Two-way ANOVA}
    \label{tab:2-way-anova}
    \begin{tabular}{|c|c|c|c|c|}
	\hline
	& sum-squared & df & F & $PR(> F)$ \\
	\hline
	\hline
    Sector & 0.06382 & 2 & 69.37 & 7.42E-24\\
    Network Architecture  & 19.1571 & 4 & 10410.81 & 3.76E-240\\
    Sector:Network & 0.18564 & 8 & 50.443 & 7.59E-45\\
    Residual & 0.9661 & 210 & &\\
    \hline
    \end{tabular}
\end{table}

We investigate each sector by performing one way ANOVA's for each sector. For each sector the factor network architecture is highly significant. We perform a multiple comparison for pairwise contrasts using a Tukey procedure. The results are presented in Tables \ref{tab:Multiple_Comparison_Rank_c3} and \ref{tab:Multiple_Comparison_Rank_c4}. We skip the detailed numerical analysis and just present the result of the analysis. In the tables Rank 1 has the best performance (percent accuracy for the test data), Rank 2 is second best performance and so on. The circled values are not statistically different. 

% \begin{table}[H]
%  \centering
%  \caption{one-way ANOVA}
%  \label{tab:one way anova}
   
%  \begin{tabular}{|c|c|c|c|c|c|}

% 	\hline

% 	Sector & & sum-squared & df & F & $PR(>F)$ \\

% 	\hline
	
% 	\hline
	 
% 	Energy & C(Model) & 7.05993 & 4 & 3941.782 & 1.39E-81 \\
% 	& Residual & 0.03134 & 70 & & \\
% 	\hline
% 	\hline
% 	Financial & C(Model) & 5.54025 & 4 & 3484.149 & 1.03E-79 \\
% 	& Residual & 0.02783 & 70 & & \\
% 	\hline
% 	\hline
% 	Healthcare & C(Model) & 6.74255 & 4 & 3151.953 & 3.36E-78 \\
% 	Residual & 0.03744 & 70 & & \\
%   	\hline
    
% \end{tabular}
 
% Table generated by Excel2LaTeX from sheet 'rank'
% Table generated by Excel2LaTeX from sheet 'rank'
\begin{table}[htbp]
  \centering
  
  \caption{Multiple Comparison results  for Case 3. Lower rank is better. Circled values are not significantly different.}
  \label{tab:Multiple_Comparison_Rank_c3}
    \begin{tabular}{|l|l|l|l|l|}
    \hline
    Rank   & \multicolumn{1}{r|}{1} & \multicolumn{1}{r|}{2} & \multicolumn{1}{r|}{3} & \multicolumn{1}{r|}{4} \\
    \hline
    Energy & \tikzmark{startup}lstm  & cnn2d\tikzmark{endup} & cnn   & mlp   \\
    Financial & \tikzmark{startup1}cnn2d & lstm\tikzmark{endup1}  & cnn   & mlp \\
    Healthcare & \tikzmark{startup2}cnn   & lstm\tikzmark{endup2}  & \tikzmark{startup3}cnn2d & mlp \tikzmark{endup3}  \\
    \hline
    \end{tabular}%
\end{table}%

\begin{table}[htbp]
  \centering
  \caption{Multiple Comparison results for Case 4. Lower rank is better. Circled values are not significantly different.}
  \label{tab:Multiple_Comparison_Rank_c4}
    \begin{tabular}{|l|l|l|l|l|}
    \hline
    Rank   & \multicolumn{1}{r|}{1} & \multicolumn{1}{r|}{2} & \multicolumn{1}{r|}{3} & \multicolumn{1}{r|}{4} \\
    \hline
    Energy &\tikzmark{startup4} lstm  & cnn2d & cnn   & mlp\tikzmark{endup4}  \\
    Financial &\tikzmark{startup5} cnn2d &\tikzmark{startup6} lstm\tikzmark{endup5}  & cnn\tikzmark{endup6}   & mlp   \\
    Healthcare & lstm  & cnn2d & cnn   & mlp  \\
    \hline
    \end{tabular}%
\begin{tikzpicture}[remember picture,overlay]
\foreach \Val in {up,up1,up2,up3,up4,up5,up6}
{
\draw[rounded corners,black,thick]
  ([shift={(-0.5\tabcolsep,-0.5ex)}]pic cs:start\Val) 
    rectangle 
  ([shift={(0.5\tabcolsep,2ex)}]pic cs:end\Val);
}
\end{tikzpicture}
\end{table}%

As expected from the two way ANOVA results for each sector we have different architectures that perform better. However, we can see that for both cases as well as all sectors the LSTM architecture is always in the top performing group. Based on these results it would appear that if one would have a single choice of network architecture the best choice is the Long Short Term Memory model.

\section{Conclusion and Future Directions}
In this work, we compared critically applications of artificial neural networks to credit risk assessment. We implement four popular neural network architectures (MLP, CNN, CNN2D, LSTM) and we used them to answer several questions relevant to finance. The results obtained point to several recommendations when applying machine learning algorithms to finance.

It is generally believed that using only a subset of relevant features produce better results for machine learning algorithms than using all the variables in the financial reports. However, in this work we found that using all the financial variables as input and allowing the neural networks to perform the feature selection in the training process works better.

All the papers applying machine learning algorithms to finance split the data randomly into a training and test set. We show that using the temporal dimension which is always present in financial data is the appropriate way to test algorithms. Specifically, we form the test set by holding out one year, and the respective year held out is the training set. Using a temporal allocation is of course how the rating is done in reality, since all quarterly reports come generally at around the same time. We show that a random allocation will produce performance results significantly higher that a proper temporal allocation so is very clear that they alter the actual algorithms performance. We believe the reason why random allocation has better performance is that some observations from the same quarter will be in the training and the test data. We think the neural network architecture picks this up.  

Finally, we investigated if there exists a neural network architecture which consistently outperform others with respect to input features, sectors and holdout set. We consider a two-way ANOVA model where the response variable is the performance result. To our knowledge this is the first proper statistical analysis of results.We conclude that for credit rating perspective a temporal network architecture such as LSTM seem to perform best. 

These recommendations are useful when assessing credit worthiness of a company not previously rated. The first time a company is rated is based primarily on historical quarterly statements. This may be the case for example when rating Twitter and LinkedIn in 2014. When the rating is updated every quarter, the assessors need to look at the change in rating and thus detecting a change in rating is perhaps a very relevant problem. We plan to investigate the performance of machine learning algorithms to detect and predict changes in credit rating in future work. 

\section*{Acknowledgements}
The authors report no conflicts of interest. The authors alone are responsible for the content and writing of the paper.
The authors are grateful to UBS for the research grant to Hanlon lab which provided partial support for this research.

\newpage
\bibliographystyle{chicago}
\bibliography{reference}

\end{document}